\documentstyle[12pt]{article}
\newcommand{\nn}{\nonumber}
\newcommand{\bd}{\begin{document}}
\newcommand{\ed}{\end{document}}
\newcommand{\bc}{\begin{center}}
\newcommand{\ec}{\end{center}}
\newcommand{\be}{\begin{eqnarray}}
\newcommand{\ee}{\end{eqnarray}}
\renewcommand{\thefootnote}{\alph{footnote}}
\newcommand{\se}{\section}
\newcommand{\sse}{\subsection}
\newcommand{\bi}{\bibitem}
\def\figcap{\section*{Figure Captions\markboth
     {FIGURECAPTIONS}{FIGURECAPTIONS}}\list
     {Figure \arabic{enumi}:\hfill}{\settowidth\labelwidth{Figure 999:}
     \leftmargin\labelwidth
     \advance\leftmargin\labelsep\usecounter{enumi}}}
\let\endfigcap\endlist \relax
\input psfig

\begin{document}
\baselineskip=0.75cm
\begin{titlepage}

 \vskip 0.5in
 \null
\begin{center}
 \vspace{.15in}

{\LARGE {\bf Probing New Physics in $B\rightarrow K^{(*)} \ell^+
\ell^-$ decays
 }}\\ \vspace{1.0cm}
\par
 \vskip 2.1em
 {\large
  \begin{tabular}[t]{c}
{\bf Chuan-Hung Chen$^a$ and C.~Q.~Geng$^{b,c}$}
\\
\\
  {\sl${}^a$ Institute of Physics, Academia Sinica}
   \\ {\sl $\ $ Taipei, Taiwan 115, Republic of China }

\\
{\sl ${}^b$Department of Physics, National Tsing Hua University}
\\  {\sl  $\ $ Hsinchu, Taiwan 300, Republic of China }
\\
{\sl ${}^c$ Theory Group, TRIUMF}
\\ {\sl 4004 Wesbrook Mall, Vancouver, B.C. V6T 2A3, Canada}
   \end{tabular}}
 \par \vskip 5.3em

\date{\today}
 {\Large\bf Abstract}
\end{center}
We study the exclusive decays of $B\to K^{(*)}\ell^+ \ell^-$ by
the results in the perturbative QCD with the heavy quark effective
theory and lattice calculations. We obtain the form factors for
the $B\to K^{(*)}$ transitions in the whole allowed region. Our
predictions on the branching ratios of $B\to K \ell^{+} \ell^{-}$,
$B\to K^{*} e^{+}e^{-}$, and $B\to K^{*} \mu^{+} \mu^{-}$ are
$0.53\pm 0.05 ^{+0.10}_{-0.07}$, $1.68\pm 0.17^{+0.14}_{-0.09}$,
and $1.34\pm 0.13 ^{+0.11}_{-0.06}\times 10^{-6}$, 
where the errors are from the quark mixing elements and 
hadronic effects, respectively.
We also find that our definitions of the
T-odd observable and the up-down asymmetry of
 the $K$ meson in $B\to K^*\ell^+ \ell^-\to K\pi\ell^+ \ell^-$
can be used to probe new physics.
\end{titlepage}

\section{Introduction}

The flavor-changing neutral current (FCNC) processes of $B\to
K^{(*)} \ell^+ \ell^-\ (\ell=e,\mu,\tau)$ are suppressed and
induced by electroweak penguin and box diagrams in the standard
model (SM) with branching ratios of ${\cal O}(10^{-7}-10^{-6})$
\cite{Ali,CK,MNS,CG-PRD}. Recently, the decay modes of $B\to K
\ell^+ \ell^-\ (\ell=e,\mu)$ have been  observed
with $Br(B\to K \ell^+ \ell^-)=(0.75^{+0.25}_{-0.21}\pm0.09)\times
10^{-6}$ \cite{Belle}
and $(0.78^{+0.24+0.11}_{-0.20-0.18})\times 10^{-6}$ \cite{BaBar,BaBar1},
at the Belle detector in the KEKB $e^+e^-$
storage ring and the BABAR detector in the PEP-II B factory
by using  $29.1\ fb^{-1}$
and $77.8\ fb^{-1}$ data simples, respectively.
%
 Experimental searches at the B-factories for $B\to K^* \ell^+
\ell^-$ are also close to the theoretical predicted ranges
\cite{Belle,BaBar,BaBar1}. 
At BABAR, an excess of events over background with estimated significance 
of $2.8\, \sigma$ has been observed and $Br(B\to K^*\ell^+
\ell^-)=(1.68^{+0.68}_{-0.58}\pm0.28)\times 10^{-6}$ has
been obtained \cite{BaBar1}.
It is clear that these FCNC rare decays
are important for not only testing the SM but probing new physics
such as supersymmetric heavy particles in SUSY models, appearing
virtually in the loop diagrams to interfere with those in the SM.

It is known that one of the main theoretical uncertainties in
studying exclusive hadron decays arises from the calculations of
matrix elements. In our previous work \cite{CG-PRD}, we 
calculated the relevant transition form factors for $B\to K (K^*)$
decays in the perturbative QCD (PQCD) approach with the
phenomenological wave functions, which are chosen by fitting with
the experimental measurements of $B\to K \pi$ and $B\to K^*
\gamma$ decays. Moreover, in order to compensate the increasing
soft gluon effects in the slow recoil region, we used trial
$q^2$-dependent wave functions instead of unknown $b$-dependent
wave functions, where $b$ is the conjugate variable of transverse
momentum of valence quark.
However,
with the trial wave functions, it is difficult to estimate the
errors in the $B\to K (K^*)$ form factors as well as the branching ratios 
and other physical observables in 
$B\to K (K^*) \ell^+ \ell^-$ decays.

To get a control of theoretical uncertainties in $B\to K
(K^*) \ell^+ \ell^-$ decays, in this paper, we first recalculate the
corresponding form factors at the large recoil region of the
momentum transfer $q^2\to 0$ with the improved PQCD approach that
includes not only  $k_{T}$ resummation, for removing end-point
singularities, but also threshold resummation, for smearing the
double logarithmic divergence arising from weak corrections
\cite{KLS-PRD}. The involved wave functions are used up to twist-3,
 derived from QCD sum rule \cite{BBKT}. Then, we
fit these results with those from the heavy quark effective theory
(HQET) and the lattice QCD (LQCD) calculations in the large $q^2$
region to get the whole $q^2$ allowed values.

To emphasis the role of new physics, we define some P and T-odd
observables. We will show that the effects of some observables are
small in the SM but can be large and measurable in models with new
physics.

The paper is organized as follows. In Sec.~II, we present the form
 factors of the $B\to K^{(*)}$ transitions in the framework of the
PQCD. In Sec.~III, we estimate the decay rates of $B\to
K^{(*)}\ell^+ \ell^-$ in the SM. We also study the polarization of
$K^*$ in $B\to K^*\ell^+ \ell^-$. In Sec.~IV, we discuss P and T
violating effects in $B\to K^*\ell^+ \ell^-\to K\pi\ell^+ \ell^-$.
We present our conclusions in Sec.~V.

\section{form factors in $B\rightarrow K^{(*)}$ transition}
To obtain the transition elements of $B\rightarrow H\left( H=K,\
K^{*}\right) $ with various weak vertices, we parametrize them in
terms of the relevant form factors as follows:\footnote{The relationships 
between various definitions of the form factors have been given in 
Appendix of Ref. \cite{CG-PRD}.}
 \begin{eqnarray}
\langle K(p_{2},\epsilon )| V_{\mu }| \bar{B} (p_{1})\rangle &=&
f_{+}(q^2)\Big\{P_{\mu}-\frac{P\cdot q }{q^2}\Big\}
+\frac{P\cdot q}{q^2}f_{0}(q^2)\,q_{\mu},  \nonumber \\
\langle K(p_{2},\epsilon )| T_{\mu\nu }q^{\nu}| \bar{B}
(p_{1})\rangle &=& {f_{T}(q^2)\over m_{B}+m_{K}}\Big\{P\cdot q\,
q_{\mu}-q^{2}P_{\mu}\Big\},
\nn \\
\langle K^{*}(p_{2},\epsilon )| V_{\mu }| \bar{B}%
(p_{1})\rangle &=&i\frac{V(q^{2})}{m_{B}+m_{K^{*}}}\varepsilon
_{\mu
\alpha \beta \rho }\epsilon ^{*\alpha }P^{\beta }q^{\rho },  \nonumber \\
\langle K^{*}(p_{2},\epsilon )| A_{\mu }| \bar{B}
(p_{1})\rangle &=&2m_{K^{*}}A_{0}(q^{2})\frac{\epsilon ^{*}\cdot q}{%
q^{2}}q_{\mu }+( m_{B}+m_{K^{*}}) A_{1}(q^{2})\Big( \epsilon
_{\mu }^{*}-\frac{\epsilon ^{*}\cdot q}{q^{2}}q_{\mu }\Big)  \nonumber \\
&&-A_{2}(q^{2})\frac{\epsilon ^{*}\cdot q}{m_{B}+m_{K^{*}}}\Big( P_{\mu }-%
\frac{P\cdot q}{q^{2}}q_{\mu }\Big) ,  \nonumber \\
\langle K^{*}(p_{2},\epsilon )| T_{\mu \nu }q^{\nu }| \bar{B}
(p_{1})\rangle &=&-iT_{1}(q^{2})\varepsilon _{\mu \alpha \beta
\rho
}\epsilon ^{*\alpha }P^{\beta }q^{\rho },  \nonumber \\
\langle K^{*}(p_{2},\epsilon )| T_{\mu \nu }^{5}q^{\nu }|
\bar{B}(p_{1})\rangle &=&T_{2}(q^{2})\Big( \epsilon _{\mu
}^{*}P\cdot q-\epsilon ^{*}\cdot qP_{\mu }\Big)
+T_{3}(q^{2})\epsilon ^{*}\cdot q\Big( q_{\mu
}-\frac{q^{2}}{P\cdot q}P_{\mu }\Big) \label{ffv}
\end{eqnarray}
where $V_{\mu }=\bar{s}\gamma _{\mu } b$, $A_{\mu }=\bar{s} \gamma
_{\mu }\gamma _{5} b$, $T_{\mu \nu }=\bar{s} i\sigma _{\mu \nu }b$, and $%
T_{\mu \nu }^{5}=\bar{s} i\sigma _{\mu \nu }\gamma _{5} b$. To
evaluate the $q^2$-dependent form factors in Eq. (\ref{ffv}), we
use two QCD methods.
One is for the large recoil region of small $q^2$ and the other
for the zero recoil region of high $q^2$.

It is known that in the large energy transfer processes, the
hadronic transition matrix elements can be calculated by the
Lepage-Brodsky (LB) \cite{LB1} formalism. However, the original LB
formalism suffers logarithmic and linear singularities in twist-2
and twist-3 wave functions 
from the end-point region
with a momentum fraction $x\to 0$, 
respectively. In order to handle these
singularities, the strategy of introducing the $k_{T}$ resummation
and threshold resummation has been proposed. It has been shown
that due the induced Sudakov factors which make the $|k_{T}|$ from
$\bar{\Lambda}$ scale to $(\bar{\Lambda}m_{B})^{1/2}$ with
$\bar{\Lambda}=m_{B}-m_b$ \cite{KLS-PRD} and $m_{b}$ being the
$b$-quark mass, the singularities do not exist in a
self-consistent improved PQCD analysis \cite{MPQCD}. Furthermore,
due to the different properties of wave functions at end-point
region, it is found that the power behaviors in $B\to K (K^*)$
form factors are the same in twist-2 and 3 wave functions
\cite{CG-NPB}. It shows the importance of twist-3 contributions.
According to Ref. \cite{CG-NPB}, we also know that other higher
twist wave functions exist extra power suppression in
$m_{0}/m_{B}$ with $m_0$ being the chiral symmetry breaking
parameter so that they are neglected in our considerations.

Recently, the applications of the PQCD approach to exclusive heavy
$B$ meson decays, such as $B\rightarrow K \pi$ \cite{KLS},
$B\rightarrow \pi \pi(KK) $ \cite{LUY,CL-PRD}, $B\rightarrow \phi
\pi(K)$ \cite{Melic,CKL-PRD}, and $B\rightarrow \rho K$
\cite{Chen-PLB} decays, have been studied and found that all of
them are consistent with the current experimental data. As known,
in these two-body charmless decays, the squared momentum transfer
is around $m^{2}_{M}$ with $M$ being a pseudoscalar or a vector
meson. That is, the PQCD can work well at the region of $q^2
\simeq m^2_{M}$. Therefore, we will apply the predicted results of
the PQCD to the large recoil region where the final outgoing meson
$M$ carries a large energy and momentum. As to the opposite region
near $q^2|_{max}$, we can use the relations among the form factor
given in the HQET \cite{iw90} by substituting the calculated
values of the form factors from the LQCD \cite{LattBK} in them to
get the remaining ones.
Once the values of the form factors at both end edges of the
allowed $q^2$ are determined, we can find the $q^2$-dependent ones
by fitting the forms
\begin{eqnarray}
F_{i}(q^2)={F_{i}(0)\over 1+\sigma_{1} s + \sigma_{2} s^2}
\label{fitf}
\end{eqnarray}
where $s=q^2/m^2_{B}$, $F_{i}(0)$ are form factors at $q^2=0$,
and $\sigma_{1,2}$ are the fitted parameters.

We now show how to obtain the form factors at large and zero
recoil:
\subsection{At large recoil}
In $B\to H \ell^+ \ell^-\ (H=K,K^*)$ decays,  the $B$ meson
momentum $p_{1}$, $H$ meson momentum $p_{2}$ and $K^*$
polarization vector
  $\epsilon$ in the $B$ meson rest frame and light-cone
coordinate are taken to be
\begin{eqnarray}
p_{1}&=& \frac{m_{B}}{\sqrt{2}}(1,1,\vec{0}_{\bot}),\ \ \ p_{2}
={\frac{m_{B}}{\sqrt{2}\eta }}(\eta
^{2},r_{H}^{2},\vec{0}_{\bot }),  \nonumber \\
\epsilon _{L} &=&{\frac{1}{\sqrt{2}r_{K^{*}}\eta }}(\eta ^{2},-r_{K^{*}}^{2},%
\vec{0}_{\bot }), \ \ \ \epsilon _{T}(\pm )
={\frac{1}{\sqrt{2}}}(0,0,1,\pm i)
\end{eqnarray}
where $\eta \simeq 1-s$ and $r_{H}=m_{H}/m_{B}$, while those for
the spectators of B and $H$ sides are expressed as
\begin{eqnarray}
k_{1}=\left(0, x_{1}{m_B \over \sqrt{2}}, \vec{k}_{1\bot}\right),
\ \ \ k_{2}=\left(x_{2}{m_B \over \sqrt{2}}\eta, 0,
\vec{k}_{2\bot}\right),
\end{eqnarray}
respectively. In our calculations, we will neglect the small
contributions from  $m_{u,d,s}$ and
 $\bar{\Lambda}$ as well as $m^{2}_{H}$ due to the on-shell
condition of the valence-quark preserved. From the
results of Ref. \cite{BBKT}, the $K^{(*)}$ meson distribution
amplitudes can be derived up to twist-3 as follows:
\begin{eqnarray}
\langle K(p)|\bar{s}(z)_{j}d(0)_{l}|0\rangle&=& -{i\over
\sqrt{2N_c}}\int^{1}_{0}dx e^{ixp\cdot z}\{\not{p}\gamma_{5}
\phi_{K}(x)+ m^0_{K}[\gamma_{5}]_{lj} \phi^{p}_{K}(x)\nonumber \\
&&+m^{0}_{K}[\gamma_{5}(\not{n_{+}}\not{n_{-}}-1)]_{lj}\phi^{t}_{K}(x)
\}, \nonumber \\
\langle K^*(p,\epsilon_{L})|\bar{s}(z)_{j}d(0)_{l}|0\rangle&=&
{1\over \sqrt{2N_c}}\int^{1}_{0}dx e^{ixp\cdot
z}\{m_{K^*}[\not{\epsilon}_{L}]_{lj}
\phi_{K^*}(x)+[\not{\epsilon}_{L}\not{p}]_{lj}\phi^{t}_{K^*}(x)\nonumber
\\ &&+m_{K^*}[I]_{lj}\phi^{s}_{K^*}(x)
\}, \nonumber \\
\langle K^*(p,\epsilon_{T})|\bar{s}(z)_{j}d(0)_{l}|0\rangle&=&
{1\over \sqrt{2N_c}}\int^{1}_{0}dx e^{ixp\cdot
z}\{m_{K^*}[\not{\epsilon}_{T}]_{lj}
\phi_{K^*}^{v}(x)+[\not{\epsilon}_{T}\not{p}]_{lj}\phi^{T}_{K^*}(x)\nonumber
\\&& +{m_{K^*}\over p\cdot n_{-}}i\varepsilon_{\mu\nu\rho\sigma}
[\gamma_{5}\gamma^{\mu}]_{lj}\epsilon^{\nu}_{T}p^{\rho}n^{\sigma}_{-}\phi^{a}_{K^*}(x)
\},\label{ss}
\end{eqnarray}
where $n_{+}=(1,0,\vec{0}_{\bot})$ and
$n_{-}=(0,1,\vec{0}_{\bot})$. In Eq. (\ref{ss}),
 $\phi_{K}(x)$, $\phi_{K^*}(x)$ and $\phi^{T}_{K^*}(x)$
are the twist-2 wave functions, and all the remaining ones
belong to the twist-3. Their explicit expressions can be found in Ref.
\cite{BBKT}.

By using the LB formalism with including $k_{T}$ and threshold
resummation, the form factors $f_{+}(q^2)$, $f_{-}(q^2)$, and
$f_{T}(q^2)$ in $B\rightarrow K$ can be written as
\begin{eqnarray}
f_{+}(q^2)&=&f_1(q^2) + f_2 (q^2)\,,
 \nonumber \\
f_{0}(q^2)&=&f_1(q^2) \Big(1+\frac{q^2}{m^2_{B}}\Big)+f_2(q^2)
\Big(1-\frac{q^2}{m^2_{B}}\Big)\,,
\label{f0}
\end{eqnarray}
where
\begin{eqnarray}
f_{1}( q^{2} ) &=&8\pi
C_{F}m_{B}^{2}r_{K}\int_{0}^{1}[dx]\int_{0}^{\infty
}b_{1}db_{1}b_{2}db_{2}\ \phi _{B}( x_{1},b_{1})  \Big[
\phi^{p}_{K}(x_2)-\phi^{t}_{K}(x_{2}) \Big]\nonumber \\
&&\times E(t^{(1)})h^{K}( x_{1},x_{2},b_{1},b_{2})\,,
 \nonumber \\
f_{2}( q^{2} ) &=&8\pi
C_{F}m_{B}^{2}\int_{0}^{1}[dx]\int_{0}^{\infty
}b_{1}db_{1}b_{2}db_{2}\ \phi
_{B}( x_{1},b_{1}) \   \nonumber \\
&&\times \Big\{ \Big[ (1+x_{2}\eta)\phi_{K}(x_{2})+2r_{K}\Big(
(\frac{1}{\eta}-x_{2})\phi^{t}_{K}(x_{2})-x_{2}\phi^{p}_{K}(x_{2})\Big)
\Big]\nonumber \\
&& \times E(t^{(1)}) h^{K}( x_{1},x_{2},b_{1},b_{2})\nonumber \\
&& +2r_{K}\phi^{p}_{K}(x_2)E(t^{(2)})h^{K}(
x_{2},x_{1},b_{2},b_{1})\Big\}\,,
\end{eqnarray}
and
\begin{eqnarray}
f_{T}( q^{2}) &=&8\pi
C_{F}m^{2}_{B}(1+r_{K})\int_{0}^{1}[dx]\int_{0}^{\infty
}b_{1}db_{1}b_{2}db_{2}\phi _{B}( x_{1},b_{1})  \nonumber \\
&&\times \Big\{ [ \phi _{K}( x_{2})-r_{K}x_{2}(\phi _{K}^{p
}(x_2)-\phi _{K}^{t }(x_2)) ] E( t^{( 1) })
h^{K}( x_{1},x_{2},b_{1},b_{2}) \nonumber \\
&& + 2r_{K}\phi _{K}^{p}( x_{2})  E( t^{( 2) }) h^{K}(
x_{2},x_{1},b_{2},b_{1}) \Big\}. \label{ft}
\end{eqnarray}
 From Eq. (\ref{f0}), we find that $f_{+}(0)=f_{0}(0)$.
The evolution factor is given by
\begin{equation}
E( t) =\alpha _{s}( t) \exp ( -S_{B}( t) -S_{K}( t) )\,,
 \label{ef}
\end{equation}
where the Sudakov exponents $S_{B(K)}$ are given in
Ref. \cite{ChenLi}. The hard functions of $h$ are written as
\begin{eqnarray}
h(x_{1},x_{2},b_{1},b_{2}) &=&S_{t}(x_{2})K_{0}(
\sqrt{x_{1}x_{2}\eta }
m_{B}b_{1})  \nonumber \\
&&\times [ \theta (b_{1}-b_{2})K_{0}( \sqrt{x_{2}\eta }
m_{B}b_{1}) I_{0}( \sqrt{x_{2}\eta }m_{B}b_{2})  \nonumber
\\ && +\theta (b_{2}-b_{1})K_{0}( \sqrt{x_{2}\eta}
m_{B}b_{2}) I_{0}( \sqrt{x_{2}\eta} m_{B}b_{1}) ] \label{dh}
\end{eqnarray}
where the threshold resummation effect is described by \cite{KLS-PRD}
\begin{eqnarray*}
S_{t}(x)={2^{1+2c}\Gamma(\frac{3}{2}+c) \over
\sqrt{\pi}\Gamma(1+c)}[x(1-x)]^{c}.
\end{eqnarray*}
 The hard scales $t^{(1,2)}$ are chosen to be
\begin{eqnarray*}
t^{( 1) } &=&\max ( \sqrt{m_{B}^{2}\eta x_{2}}
,1/b_{1},1/b_{2}) \,,  \nonumber \\
t^{( 2) } &=&\max ( \sqrt{m_{B}^{2}\eta x_{1}},1/b_{1},1/b_{2}).
\label{tscale}
\end{eqnarray*}
In Ref. \cite{CG-NPB}, we have given detailed discussions and
expressions for all
 form factors in the $B\rightarrow K^*$ transition based on the PQCD
in the large recoil region. We emphasize that in the PQCD approach
there is an identity at $q^2=0$, given by \cite{CG-NPB}
\begin{eqnarray*}
A_{2}(0) &=&\left( 1+r_{K^{*}}\right)
^{2}A_{1}(0)-2r_{K^{*}}\left( 1+r_{K^{*}}\right) A_{0}(0)\,,
 \label{idff}\\
\end{eqnarray*}
which is consistent with the leading order model-independent relation
\cite{Ali,BF,Charles,Burdman,MS,LF}
\begin{eqnarray*}
A_{2}(0) &=&{1+r_{K^{*}}\over 1-r_{K^*}} A_{1}(0)-{2r_{K^{*}}
\over 1-r_{K^*}}  A_{0}(0).
\end{eqnarray*}

In our numerical calculations, we use
 $f_B=0.19$
GeV, $f_{K^*}=0.21$ GeV, $f^{T}_{K^*}=0.17$ GeV, $m_{B}=5.28$ GeV,
$m_{K^*}=0.892$ GeV, and $c=0.3\ (0.4)$ for $B\rightarrow K\ (K^*)$,
and we take the $B$ meson wave function
as \cite{ChenLi,KLS-PRD}
\begin{equation}
\phi _{B}(x,b)=N_{B}x^{2}(1-x)^{2}\exp \left[ -\frac{1}{2}\left( \frac{xM_{B}%
}{\omega _{B}}\right) ^{2}-\frac{\omega _{B}^{2}b^{2}}{2}\right] ,
\label{bwf}
\end{equation}
where $\omega _{B}$ is the shape parameter \cite{BW} and $N_{B}$
is determined by the normalization of the wave function, given
by
\begin{eqnarray*}
\int_{0}^{1}dx\phi_{B}(x,0)={f_B \over 2\sqrt{2N_{c}}}.
\end{eqnarray*}
Since the shape parameter $\omega_{B}$ and chiral symmetry
breaking parameter $m^0_{K}$ are free in the PQCD, in order to
estimate the uncertainties, we choose (I) $\omega_{B}=0.40$ and
(II) $\omega_{B}=0.42$ for $B\rightarrow K^*$ and (I)
$\omega_{B}=0.40$ and $m^0_{K}=1.7$ and (II) $\omega_{B}=0.42$ and
$m^0_{K}=1.5$ for $B\rightarrow K$ as the upper and lower bounds,
respectively. From these values, we obtain the form factors of
$B\rightarrow K$ at $q^2=0$ as: (I) $f_{+}=0.354$ and
$f_{T}=0.250$ and (II) $f_{+}=0.303$ and $f_{T}=0.220$, while
those for the $B\rightarrow K^*$ ones are shown in Table
{\ref{vff}}. It is known that at the small $q^2$ region there
exists a large energy effective theory (LEET) so that all form
factors for heavy-to-light decays can be described by few
independent functions \cite{Charles}. In Table \ref{vff}, we also
show the results found by combining the LEET with the experimental
data of $B\rightarrow K^* \gamma$ \cite{Burdman}. 
We note that our results for the form factors are different from those in 
Ref. \cite{CG-PRD}, which shall be referred as (III).
We remark that we have neglected nonlocal contributions from the 
four-quark operators. These effects can contribute in the $5-10\%$ range 
to the form factors \cite{BFS}. 
\begin{table}
\caption{ Form factors for
$B\rightarrow K^*$ in the LEET and PQCD with
(I) $\omega_{B}=0.40$ and (II) $\omega_{B}=0.42$. }
\label{vff}
\begin{center}
\begin{tabular}{lllllll}
\hline
& \multicolumn{1}{c}{$V(0)$} & $A_{0}( 0)$ &
\multicolumn{1}{c}{$A_{1}(0)$} & $ A_{2}(0)$
 & \multicolumn{1}{c}{$T_{1}(0)$} & $T_{3}(0)$ \\
\hline LEET\cite{Burdman} & \multicolumn{1}{c}{$0.36\pm 0.04$} &
\multicolumn{1}{c}{} & \multicolumn{1}{c}{$0.27\pm 0.03$} &
\multicolumn{1}{c}{} &
\multicolumn{1}{c}{$0.31\pm 0.02$} & \multicolumn{1}{c}{} \\
\hline PQCD (I) & \multicolumn{1}{c}{$0.355$} &
\multicolumn{1}{c}{$0.407$} & \multicolumn{1}{c}{$0.266$} &
\multicolumn{1}{c}{$0.202$} & \multicolumn{1}{c}{$0.315$} &
\multicolumn{1}{c}{$0.207$} \\
\multicolumn{1}{r}{(II)} & \multicolumn{1}{c}{$0.332$} & \multicolumn{1}{c}{$%
0.381$} & \multicolumn{1}{c}{$0.248$} &
\multicolumn{1}{c}{$0.189$}
& \multicolumn{1}{c}{$0.294$} & \multicolumn{1}{c}{$0.193$} \\
\hline
\end{tabular}
\end{center}
\end{table}
\subsection{At zero recoil}
\subsubsection{Form factors in $B\rightarrow K$}
According to the analysis of Ref. \cite{iw90},  under the heavy
quark symmetry, the form factor of $f_{T}(q^2)$ in $B\rightarrow K$
 can be written in terms of two independent form factors as
\begin{eqnarray}
f_{T}(q^2)&=&\frac{m_B+m_P}{2m_{B}}(f_{+}(q^2)-f_{-}(q^2))
\label{hqetft}
\end{eqnarray}
where $f_{-}(q^2)=(m^2_{B}-m^2_{K})(f_{0}(q^2)-f_{+}(q^2))/
q^{2}$.
Since there is no complete lattice calculations on
the form factors in $B\rightarrow K$ in the literature yet, the
strategy to obtain $f_{+}$ and $f_{-}$ is that we utilize the
relationships, which connect the form factors between
$B\rightarrow K $ and $D\rightarrow K$ with the  HQET,
given by \cite{iw90}
\begin{eqnarray}
f_+^B(v\cdot
p)&=&\frac{1}{2}(\frac{m_B}{m_D})^{1/2}\Big({\alpha_{s}(m_{b})
\over \alpha_{s}(m_{c})}\Big)^{-(6/25)} \Big[(1+\frac{m_D}{m_B})
f_+^D(v\cdot p)-(1-\frac{m_D}{m_B})f_-^D(v\cdot p)\Big],
\nonumber\\
f_-^B(v\cdot
p)&=&\frac{1}{2}(\frac{m_B}{m_D})^{1/2}\Big({\alpha_{s}(m_{b})
\over \alpha_{s}(m_{c})}\Big)^{-(6/25)}
\Big[(1+\frac{m_D}{m_B})f_-^D(v\cdot p)
-(1-\frac{m_D}{m_B})f_+^D(v\cdot p)\Big], \label{hqetbk}
\end{eqnarray}
where $\Big({\alpha_{s}(m_{b})/ \alpha_{s}(m_{c})}\Big)^{-(6/25)}$
is the relevant renormalization  factor. We note that the form
factors $f^{B}_{+(-)}(v\cdot p)$ are evaluated at the same value
of $v\cdot p$ as  $f^{D}_{+(-)}(v\cdot p)$. In order to get the
values at zero recoil, we adopt the lattice results in
$D\rightarrow K$ \cite{LattDK}, where the predicted BR for
$D\rightarrow K \ell \nu$ is consistent with the measurement
\cite{PDG}. From  Eqs. (\ref{hqetft}) and (\ref{hqetbk}), we find
the form factors shown in Table \ref{lattice} in the large $q^2$
region. In Table \ref{pqcdhqetlattp}, we show the form factors in
$B\rightarrow K$ at $q^2=0$ and fitted parameters of
$\sigma_{1,2}$ in Eq. (\ref{fitf}).

\begin{table}
\caption{ Form factors in $B\rightarrow K$ from  the HQET
and lattice calculations at some large values of  $q^2$.}
\begin{center}
\label{lattice}
\begin{tabular}{llll} %
\hline $q^2$ GeV$^{2}$ & \multicolumn{1}{c}{$f_{+}(q^2)$} &
\multicolumn{1}{c}{$f_{0}(q^2)$} &
\multicolumn{1}{c}{$f_{T}( q^2)$} \\
\hline 21  & \multicolumn{1}{c}{$1.61 \pm 0.11$} &
\multicolumn{1}{c}{$0.61\pm 0.03$} & \multicolumn{1}{c}{$ 1.54\pm 0.12$}  \\
\hline 22  & \multicolumn{1}{c}{$ 2.05 \pm 0.12$} &
\multicolumn{1}{c}{$0.67 \pm 0.03$} & \multicolumn{1}{c}{$1.99\pm 0.13$}  \\
\hline
\end{tabular}
\end{center}
\end{table}

\begin{table}
\caption{ Form factors in $B\rightarrow K$ at $q^2=0$ and
fitted parameters of  $\sigma_{1,2}$ in Eq. (\ref{fitf}).}
\label{pqcdhqetlattp}
\begin{center}
\begin{tabular}{llll} %
\hline  & \multicolumn{1}{c}{$f_{+}(q^2)$} & \multicolumn{1}{c}{$
f_{0}(q^2)$}& \multicolumn{1}{c}{$f_{T}(q^2)$}  \\
\hline (I) $q^2=0$  & \multicolumn{1}{c}{$0.354$} &
\multicolumn{1}{c}{$0.354$} & \multicolumn{1}{c}{$0.25$} \\
$\ \ \ \ \sigma_{1}$ & \multicolumn{1}{c}{$-1.246$} &
\multicolumn{1}{c}{$-0.297$} & \multicolumn{1}{c}{$-1.570$} \\
$\ \ \ \ \sigma_{2}$ & \multicolumn{1}{c}{$0.251$}&
\multicolumn{1}{c}{$-0.400$}  & \multicolumn{1}{c}{$0.584$} \\
\hline (II) $q^2=0$  & \multicolumn{1}{c}{$0.303$} &
\multicolumn{1}{c}{$0.303$} & \multicolumn{1}{c}{$0.220$} \\
$\ \ \ \ \sigma_{1}$  & \multicolumn{1}{c}{$-1.229$} &
\multicolumn{1}{c}{$-1.212$} & \multicolumn{1}{c}{$-1.496$} \\
$\ \ \ \ \sigma_{2}$ & \multicolumn{1}{c}{$0.219$}  &
\multicolumn{1}{c}{$0.755$} & \multicolumn{1}{c}{$0.492$} \\
\hline
\end{tabular}
\end{center}
\end{table}

\subsubsection{ Form factors in $B\rightarrow K^*$}
We calculate the form factors in $B\rightarrow K^*$ by using the same
technique as that in $B\rightarrow K$.
Following the HQET, we have the relations given by \cite{iw90,bd91}
\begin{eqnarray}
  T_1(q^2) &=& \frac{m_B^2+q^2-m_V^2}{2m_B}\frac{V(q^2)}
   {m_B+m_V}+\frac{m_B+m_V}{2m_B} A_1(q^2),
\nonumber\\
T_1(q^2)-T_2(q^2) &=&\frac{q^2}{m_B^2-m_V^2}\Big[
   \frac{3m_B^2-q^2+m_V^2}{2m_B}\frac{V(q^2)}
   {m_B+m_V}-\frac{m_B+m_V}{2m_B} A_1(q^2) \Big],
\label{rel_3} \nonumber\\
  T_3(q^2) &=& \frac{m_B^2-q^2+3m_V^2}{2m_B} \frac{V(q^2)}{m_B+m_V}+
\frac{m_B^2-m_V^2}{m_B q^2} m_V A_0(q^2), \nonumber\\
&\hspace*{-1cm}-&\hspace*{-0.5cm}\frac{m_B^2+q^2-m_V^2}{2m_B q^2}
  \Big[(m_B+m_V)A_1(q^2)-(m_B-m_V)A_2(q^2)\Big] \label{hqetvff}.
\end{eqnarray}
We remark that the above identities are valid only for $q^2$ close
to the zero recoil region where the HQET is reliable. 
 We note that the relations in  Eq.
(\ref{hqetvff}) are not complete as they mix terms of different order in 
$1/m_b$ in the zero recoil region \cite{BB}. The correct treatment has 
been given recently in Ref. \cite{GP}.
 From Eq.
(\ref{hqetvff}), we find that $V(q^2)$, $A_{1}(q^2)$, and
$T_{3}(q^2)$ can be determined once $T_{1,2}(q^2)$ and
$A_{0,2}(q^2)$ are fixed. In Table \ref{Vlattice}, we display the
form factors in $B\rightarrow K^*$, where we have used Eq.
(\ref{hqetvff}) in the HQET and the lattice QCD results
\cite{LattBK} of $T_{1,2}(q^2)$ and $A_{0,2}(q^2)$, which  have
been demonstrated to be consistent with the measurement in
$B\rightarrow K^* \gamma$.
 We remark that
$A_{0}(q^2)=T_{1}(q^2)$ has been taken in the lattice calculations.
The form factors in $B\rightarrow K^*$ at $q^2=0$ and
fitted parameters of  $\sigma_{1,2}$ in Eq. (\ref{fitf}) are shown in
 Table \ref{pqcdhqetlattv}.

\begin{table}[hpt]
\caption{ Form factors for $B\rightarrow K^*$ from  the lattice
results of $T_{1,2}(q^2)$ and $A_{0,2}(q^2)$ with the HQET at some values 
of higher $q^2$.}
 \label{Vlattice}
\begin{center}
\begin{tabular}{lllllll} %
\hline
$q^2$ GeV$^{2}$ & \multicolumn{1}{c}{$V( q^2)$} &
\multicolumn{1}{c}{$A_{1}(q^2)$} &
\multicolumn{1}{c}{$A_{2}(q^2)$} &
\multicolumn{1}{c}{$T_{1}(q^2)$} &
\multicolumn{1}{c}{$T_{2}(q^2)$} &
\multicolumn{1}{c}{$T_{3}(q^2)$} \\
\hline 16  & \multicolumn{1}{c}{$1.33 \pm 0.05$} &
\multicolumn{1}{c}{$0.44 \pm 0.02$} & \multicolumn{1}{c}{$ 0.67
\pm 0.03$} & \multicolumn{1}{c}{$1.14 \pm 0.04$}&
\multicolumn{1}{c}{$0.47 \pm 0.02$} &  \multicolumn{1}{c}{$ 0.67 \pm 0.03$}\\
\hline 17  & \multicolumn{1}{c}{$ 1.50\pm 0.05$}  &
\multicolumn{1}{c}{$ 0.46\pm 0.01$}& \multicolumn{1}{c}{$0.72 \pm
0.02$}& \multicolumn{1}{c}{$1.28 \pm 0.04 $}&
\multicolumn{1}{c}{$0.48 \pm 0.01$}& \multicolumn{1}{c}{$0.73 \pm 0.02$}  \\
\hline 19  & \multicolumn{1}{c}{$ 1.94\pm 0.03$}  &
\multicolumn{1}{c}{$0.49 \pm 0.01$}& \multicolumn{1}{c}{$0.82 \pm
0.01 $}&\multicolumn{1}{c}{$ 1.66\pm 0.02 $}&
\multicolumn{1}{c}{$0.49 \pm 0.01$}& \multicolumn{1}{c}{$0.86 \pm 0.01$}  \\
\hline
\end{tabular}
\end{center}
\end{table}

\begin{table}[hpt]
\caption{ Form factors in $B\rightarrow K^*$ at $q^2=0$ and
fitted parameters of  $\sigma_{1,2}$ in Eq. (\ref{fitf}).}
\label{pqcdhqetlattv}
\begin{center}
\begin{tabular}{llllllll} %
\hline  & \multicolumn{1}{c}{$V(q^2)$} &
\multicolumn{1}{c}{$A_{0}(q^2)$} & \multicolumn{1}{c}{$A_{1}(q^2)$} &
\multicolumn{1}{c}{$A_{2}(q^2)$} &
\multicolumn{1}{c}{$T_{1}(q^2)$} &
\multicolumn{1}{c}{$T_{2}(q^2)$} &
\multicolumn{1}{c}{$T_{3}(q^2)$} \\
\hline (I) $q^2=0$  & \multicolumn{1}{c}{$0.355$} &
\multicolumn{1}{c}{$0.407$} & \multicolumn{1}{c}{$0.266$} &
\multicolumn{1}{c}{$ 0.202$} & \multicolumn{1}{c}{$0.315$}&
\multicolumn{1}{c}{$0.315$} &  \multicolumn{1}{c}{$ 0.207$}\\
$\ \ \ \ \sigma_{1}$ & \multicolumn{1}{c}{$-1.802$} &
\multicolumn{1}{c}{$-1.282$} & \multicolumn{1}{c}{$-1.034$} &
\multicolumn{1}{c}{$ -1.906$} & \multicolumn{1}{c}{$-1.749$}&
\multicolumn{1}{c}{$-0.975$} &  \multicolumn{1}{c}{$-1.777$}\\
$\ \ \ \ \sigma_{2}$ & \multicolumn{1}{c}{$0.879$}&
\multicolumn{1}{c}{$0.249$}  & \multicolumn{1}{c}{$0.514$} &
\multicolumn{1}{c}{$ 1.168$} & \multicolumn{1}{c}{$0.816$}&
\multicolumn{1}{c}{$0.632$} &  \multicolumn{1}{c}{$0.964$}\\
\hline (II) $q^2=0$  & \multicolumn{1}{c}{$0.332$} &
\multicolumn{1}{c}{$0.381$} & \multicolumn{1}{c}{$0.248$} &
\multicolumn{1}{c}{$ 0.189$} & \multicolumn{1}{c}{$0.294$}&
\multicolumn{1}{c}{$0.294$} &  \multicolumn{1}{c}{$ 0.193$}\\
$\ \ \ \ \sigma_{1}$  & \multicolumn{1}{c}{$-1.721$} &
\multicolumn{1}{c}{$-1.228$} & \multicolumn{1}{c}{$-0.829$}&
\multicolumn{1}{c}{$-1.801$}& \multicolumn{1}{c}{$-1.671 $}&
\multicolumn{1}{c}{$-0.721$}& \multicolumn{1}{c}{$-1.677$}  \\
$\ \ \ \ \sigma_{2}$ & \multicolumn{1}{c}{$0.744$}  &
\multicolumn{1}{c}{$0.148$} & \multicolumn{1}{c}{$0.166$}&
\multicolumn{1}{c}{$0.993$}&\multicolumn{1}{c}{$0.684$}&
\multicolumn{1}{c}{$0.202$}& \multicolumn{1}{c}{$0.794$}  \\
\hline
\end{tabular}
\end{center}
\end{table}

\section{Differential Decay Rates and
polarizations}

 From the definitions of form factors in Eq. (\ref{ffv}),
in the SM the transition amplitudes for $B\rightarrow
K^{(*)}\ell^{+}\ell^{-}$ ($\ell=e, \mu$) can be written as
\begin{equation}
{\cal M}_{K}=\frac{G_{F}\alpha \lambda _{t}}{2\sqrt{2}\pi }\Big\{
 \Big[ C^{eff}_{9}(\mu)f_{+}(q^{2}) +
2m_{b}C_{7}(\mu){f_{T}( q^{2})\over m_{B}+m_{K}} \Big] P_{\mu } \
\bar{\ell}\gamma^{\mu} \ell +C_{10}f_{+}(q^2)P_{\mu
}\bar{\ell}\gamma^{\mu}\gamma_5 \ell\Big\} \label{ampk}
\end{equation}
and
\begin{equation}
{\cal M}_{K^{*}}^{(\lambda )}=\frac{G_{F}\alpha \lambda _{t}}{2\sqrt{2}%
\pi }\left\{ {\cal M}_{1\mu }^{(\lambda )}\bar{\ell}\gamma^{\mu}
\ell+{\cal M}_{2\mu }^{(\lambda )}\bar{\ell}\gamma^{\mu}\gamma_5
\ell \right\} \label{ampk*}
\end{equation}
with
\begin{eqnarray}
{\cal M}_{i\mu }^{(\lambda )} &=&i\xi_{1}\varepsilon _{\mu \nu
\alpha \beta }\epsilon ^{*\nu }(\lambda )P^{\alpha }q^{\beta
}+\xi_{2}\epsilon _{\mu }^{*}(\lambda )+\xi_{3}\epsilon ^{*}\cdot
qP_{\mu },
\label{ampk1}
\end{eqnarray}
where we have set $m_{\ell}=0\ (\ell=e,\,\mu)$ ,
$\lambda_t=V_{tb}V_{ts}^*\simeq 0.041\pm 0.002$ \cite{Buras0},
$C_i^{(eff)}$ are the Wilson coefficients (WCs) and their
expressions can be found in Refs. \cite{CG-PRD,Buras}, and $i=1\
(2)$ for $\xi_j=h_j\ (g_j)$ with $j=1,2,3$, defined in Appendix.

 For the decays of $B\to K \ell^{+}\ell^{-}$,
since the $K$ meson is a pseudo-scalar, the only interesting physical
observables are the decay rates themselves, whereas
  other observables such as the lepton polarizations, discussed
in detail in Refs.  \cite{Ali,CK,MNS,CG-PRD}, are hard to be measured
by experiments. The differential decay rates for $B\to K
\ell^+ \ell^-$ are given by \cite{CG-PRD}
\begin{equation}
\frac{d\Gamma _{K}( s) }{ds}=\frac{G_{F}^{2}\alpha^{2}| \lambda
_{t}| ^{2}m_{B}^{5}}{3\times 2^{9}\pi ^{5}}(1-s)^{3/2} \Big[ \Big|
C_{9}^{eff}(\mu) f_{+}(q^2) +2m_{b}C_{7}(\mu)f_{T}( q^{2}) \Big|
^{2}+\Big| C_{10}f_{+}( q^{2}) \Big|^{2} \Big]
 \label{dr}
\end{equation}
where $s=q^2/M_B^2$.
For $B\rightarrow K^* \ell^{+} \ell^{-}$,
however,
one can study not only
the decay rates and lepton polarizations but also
physical observables related to the $K^*$ polarization,
including longitudinal and transverse polarizations of $K^*$ and P
and T-odd
observables.
To analyze  the polarization of $K^*$,
we have to consider the decay chain
$B\rightarrow K^*\ell^+ \ell^-\rightarrow K \pi\ell^+ \ell^-$,
and we
choose the $K^*$ helicities as  $\epsilon
(0)=(|\vec{p}_{K^{*}}|,0,0,E_{K^{*}})/M_{K^{*}}$
and  $\epsilon (\pm)=(0,1,\pm i,0)/\sqrt{2}$, the positron lepton momentum
$p_{l^{+}}=\sqrt{q^{2}}(1,\sin \theta _{l},0,\cos \theta _{l})/2$
with $E_{K^{*}}=(M_{B}^{2}-M_{K^{*}}^{2}-q^{2})/2\sqrt{q^{2}}$ and
$|\vec{p}_{K^{*}}|=\sqrt{E_{K^{*}}^{2}-M_{K^{*}}^{2}}$ in the
$q^{2}$ rest frame, and the $K$ momentum  $p_{K}=(1,\sin \theta
_{K}\cos \phi ,\sin \theta _{K}\sin \phi ,\cos \theta
_{K})M_{K^{*}}/2$ in the $K^{*}$ rest frame where $\phi $ denotes
the relative angle of decaying plane between $K\pi $ and
$l^{+}l^{-}$. From Eq. (\ref{ampk*}), the
differential decay rates of
$B\rightarrow K^*\ell^+ \ell^-\rightarrow K \pi\ell^+ \ell^-$
 as functions of angles $\theta _{K} $ and
$\phi $ are found to be
\begin{eqnarray}
\frac{d\Gamma }{d\cos \theta _{K}d\phi dq^{2}} &=&\frac{
G_{F}^{2}\alpha^{2}|\lambda _{t}| ^{2}| \vec{p} |
}{2^{14}\pi ^{6}m_{B}^{2}}Br(K^{*}\rightarrow K\pi ) \nonumber \\
&&\times \Big\{16\cos ^{2}\theta _{K}\sum_{i=1,2}|
{\cal M}_{i}^{0}| ^{2} \nonumber \\
&& +8\sin ^{2}\theta _{K}\sum_{i=1,2}\Big( | {\cal
M}_{i}^{+}|^{2}+| {\cal M}_{i}^{-}|^{2}\Big) \nonumber \\
&&-8\sin ^{2}\theta _{K}\Big[ \cos 2\phi \sum_{i=1,2}{Re }{\cal
M}_{i}^{+}{\cal M}_{i}^{-*} +\sin 2\phi \sum_{i=1,2}{Im}
 {\cal M}_{i}^{+}{\cal M}_{i}^{-*}\Big]  \nonumber \\
&&+3\pi\sin 2\theta _{K}\Big[ \cos \phi \Big( {Re}{\cal M}
_{1}^{0}({\cal M}_{2}^{+*}-{\cal M}_{2}^{-*})+{Re}({\cal
M}_{1}^{+}-{\cal M}
_{1}^{-}){\cal M}_{2}^{0*}\Big)   \nonumber \\
&&+\sin \phi \Big( {Im}{\cal M}_{1}^{0}({\cal M}_{2}^{+*}+ {\cal
M}_{2}^{-*})-{Im}({\cal M}_{1}^{+}+{\cal M}_{1}^{-}){\cal
M}_{2}^{0*}\Big) \Big] \Big\}  \label{difangle}
\end{eqnarray}
with
\begin{eqnarray}
|\vec{p}|&=&\sqrt{ E^{\prime 2}- m^{2}_{K^*} },\ \ \  E^{\prime
}={m^{2}_{B}+m^{2}_{K^*}-q^{2} \over 2m_{B}}\,,
\nonumber \\
{\cal M}_{a}^{0} &=&\sqrt{q^{2}}\Big( \frac{E_{K^{*}}}{m_{K^{*}}}
\xi_{2}+2 \sqrt{q^{2}}\frac{| p_{K^{*}}|^{2} }{
m_{K^{*}}}\xi_{3}\Big)\,,
  \nonumber \\
{\cal M}_{a}^{\pm } &=&\sqrt{q^{2}}( \pm 2| p_{K^{*}}| \sqrt{
q^{2}}\xi_{1}+\xi_{2})\,,
  \label{M}
\end{eqnarray}
where $a=1(2)$.
We note that other discussions for
 various polarizations can be in
 Refs. \cite{Kim,Kruger,Grossman}.
 From  Eq. (\ref{difangle}), by integrating all angles
 we obtain
\begin{eqnarray}
\frac{d\Gamma_{K^*}(s)}{ds}
&=&\frac{G_{F}^{2}\alpha^{2}| \lambda_{t}| ^{2}| \vec{p}|
}{3\times 2^{8}\pi ^{5}}
 \sum_{\lambda=+,0,- }\sum_{i=1,2}|{\cal M}_{i}^{\lambda }|^{2}\,,
 \label{difrate}
\end{eqnarray}
where we have used that $Br(K^*\to K\pi)=1$ \cite{PDG}.

The components ${\cal M}_{a}^{0}$ and ${\cal M}_{a}^{\pm }$ in Eq.
(\ref{M}) clearly denote the longitudinal and transverse
polarizations, which can be extracted by integrating out the angle
$\phi$ dependence in Eq. (\ref{difangle}), respectively, and
explicitly we have that
\begin{eqnarray}
{d\Gamma \over dq^2 d\cos\theta_{K}}&=&\frac{
G_{F}^{2}\alpha^{2}|\lambda _{t}| ^{2}| \vec{p} | }{2^{10}\pi
^{5}m_{B}^{2}}Br(K^{*}\rightarrow K\pi ) \Big\{2\cos ^{2}\theta
_{K}\sum_{i=1,2}|
{\cal M}_{i}^{0}| ^{2} \nonumber \\
&& +\sin ^{2}\theta _{K}\sum_{i=1,2}\Big( | {\cal
M}_{i}^{+}|^{2}+| {\cal M}_{i}^{-}|^{2}\Big)\Big\},
\label{ltp}
\end{eqnarray}
 From Eqs. (\ref{difrate}) and (\ref{ltp}),
we may define
\begin{eqnarray}
{\cal P}_{L}(q^2)&=& {\sum_{i=1,2}| {\cal M}_{i}^{0}| ^{2} \over
\sum_{\lambda=+,0,- }\sum_{i=1,2}|{\cal
M}_{i}^{\lambda }|^{2}}, \label{pl}\\
{\cal P}_{T}(q^2)&=& {\sum_{i=1,2}| {\cal M}_{i}^{+}| ^{2}+| {\cal
M}_{i}^{-}| ^{2} \over \sum_{\lambda=+,0,- }\sum_{i}|{\cal
M}_{i}^{\lambda }|^{2} }, \label{pt}
\end{eqnarray}
as the normalized longitudinal and transverse parts with their
ratio being
\begin{eqnarray}
\xi(q^2)={{\cal P}_{T}(q^2) \over {\cal P}_{L}(q^2)}={
 \sum_{i=1,2}| {\cal
M}_{i}^{+}| ^{2}+| {\cal M}_{i}^{-}| ^{2} \over \sum_{i=1,2}|
{\cal M}_{i}^{0}| ^{2} }\,.
\label{xi}
\end{eqnarray}

\begin{figure}[tbp]
\vspace{0cm} \centerline{ \psfig{figure=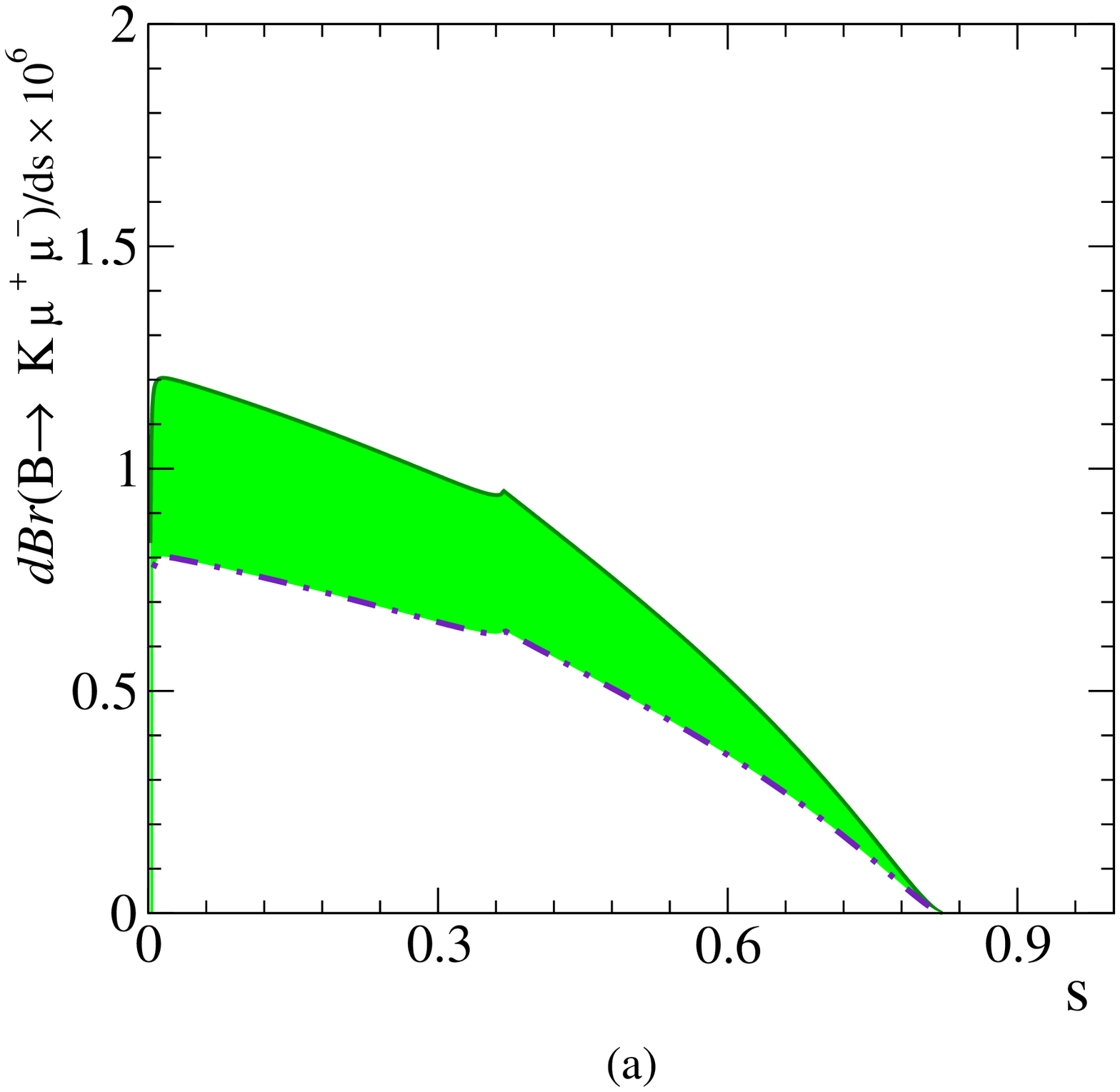,height=3.in } $\
\ $ \psfig{figure=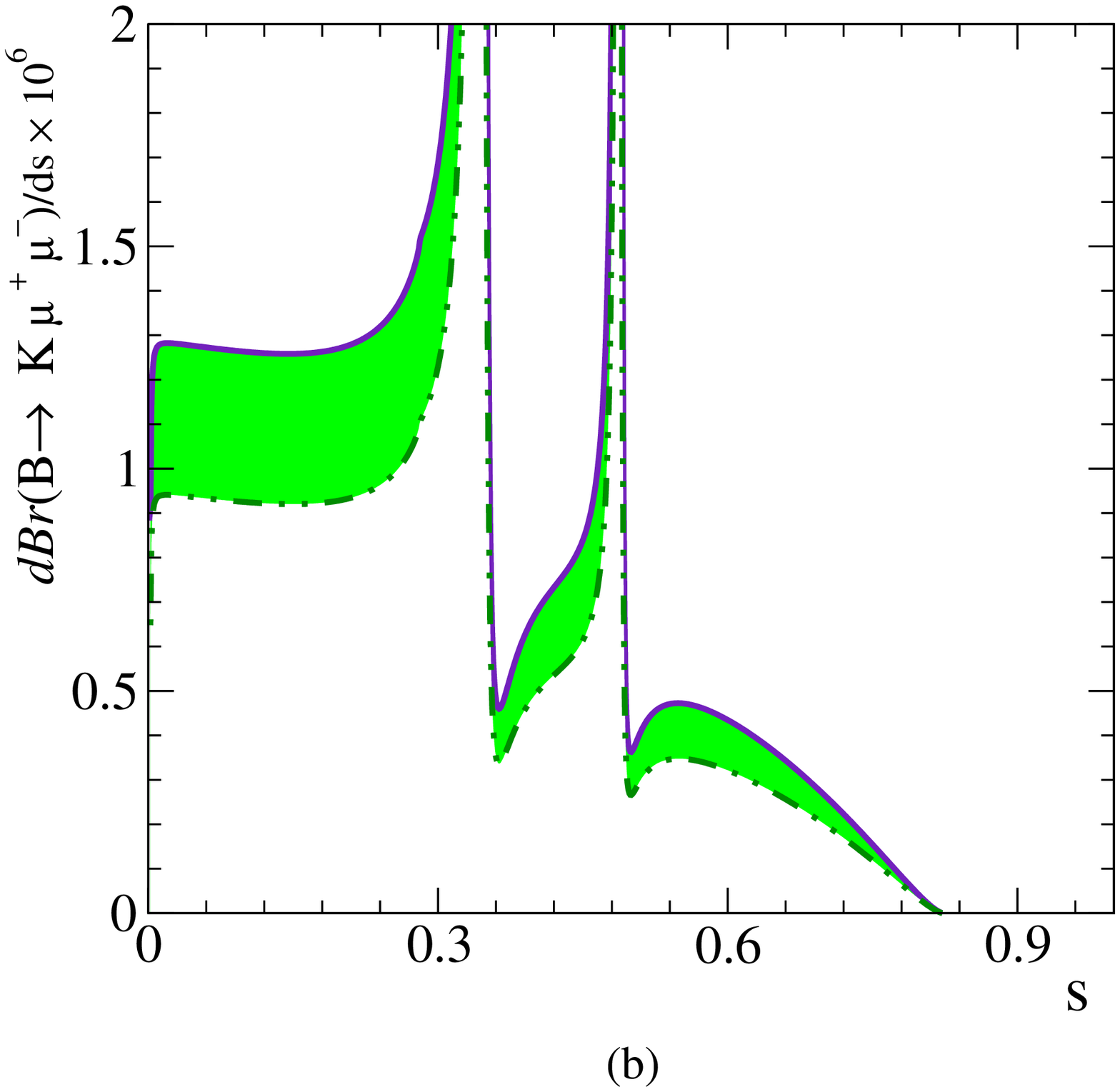,height=3.in } }
\caption{Differential decay BR of $B\rightarrow K \mu^+ \mu^-$
with (a) and (b) representing the results with and without
resonant effect, respectively. The solid (dash-dotted) curve
stands for the upper (lower) bound with the allowed region being
shaded. }\label{difratek}
\end{figure}

\begin{figure}[tbp]
\vspace{2cm} \centerline{ \psfig{figure=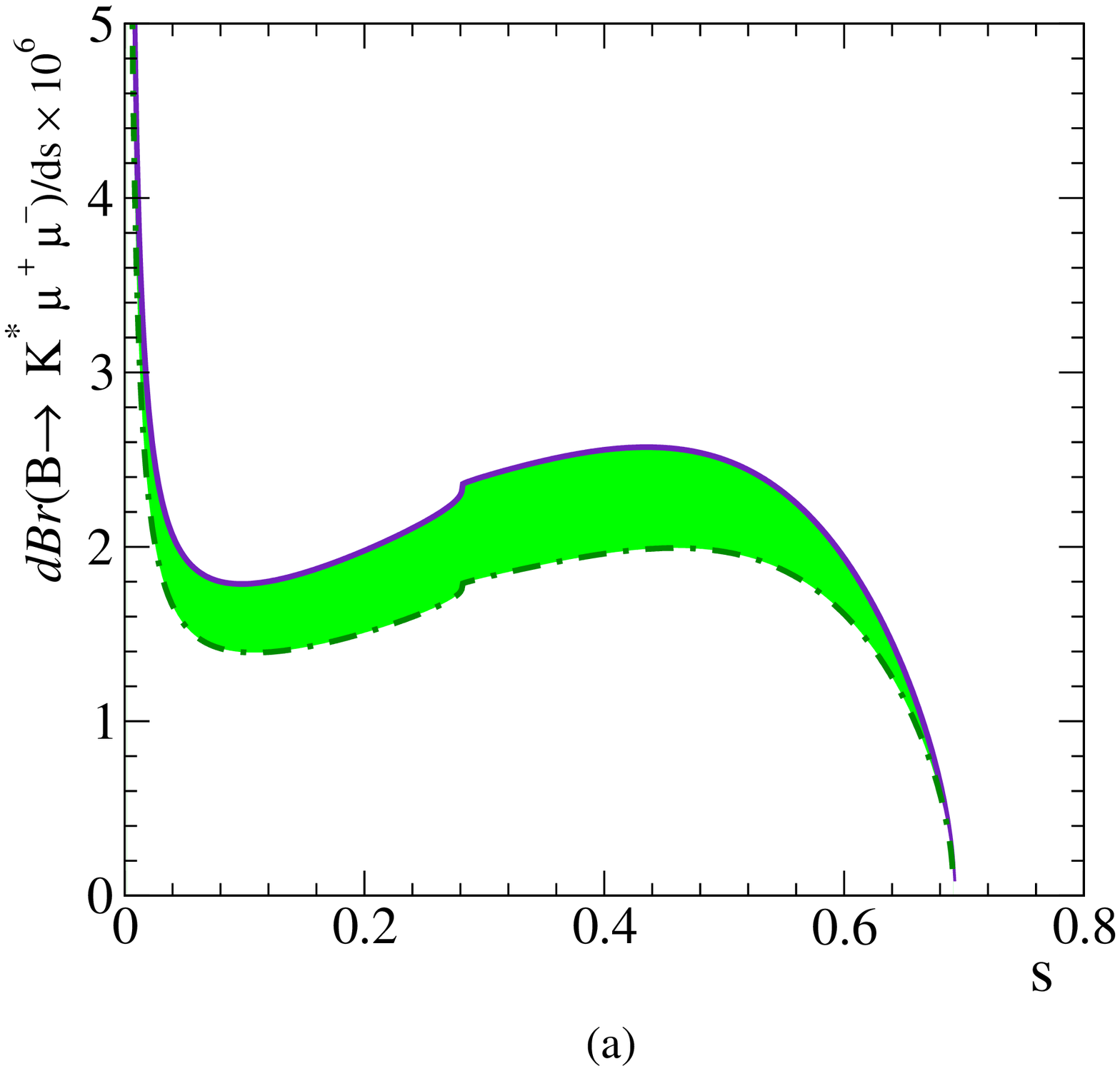,height=3.in }
$\ \ $ \psfig{figure=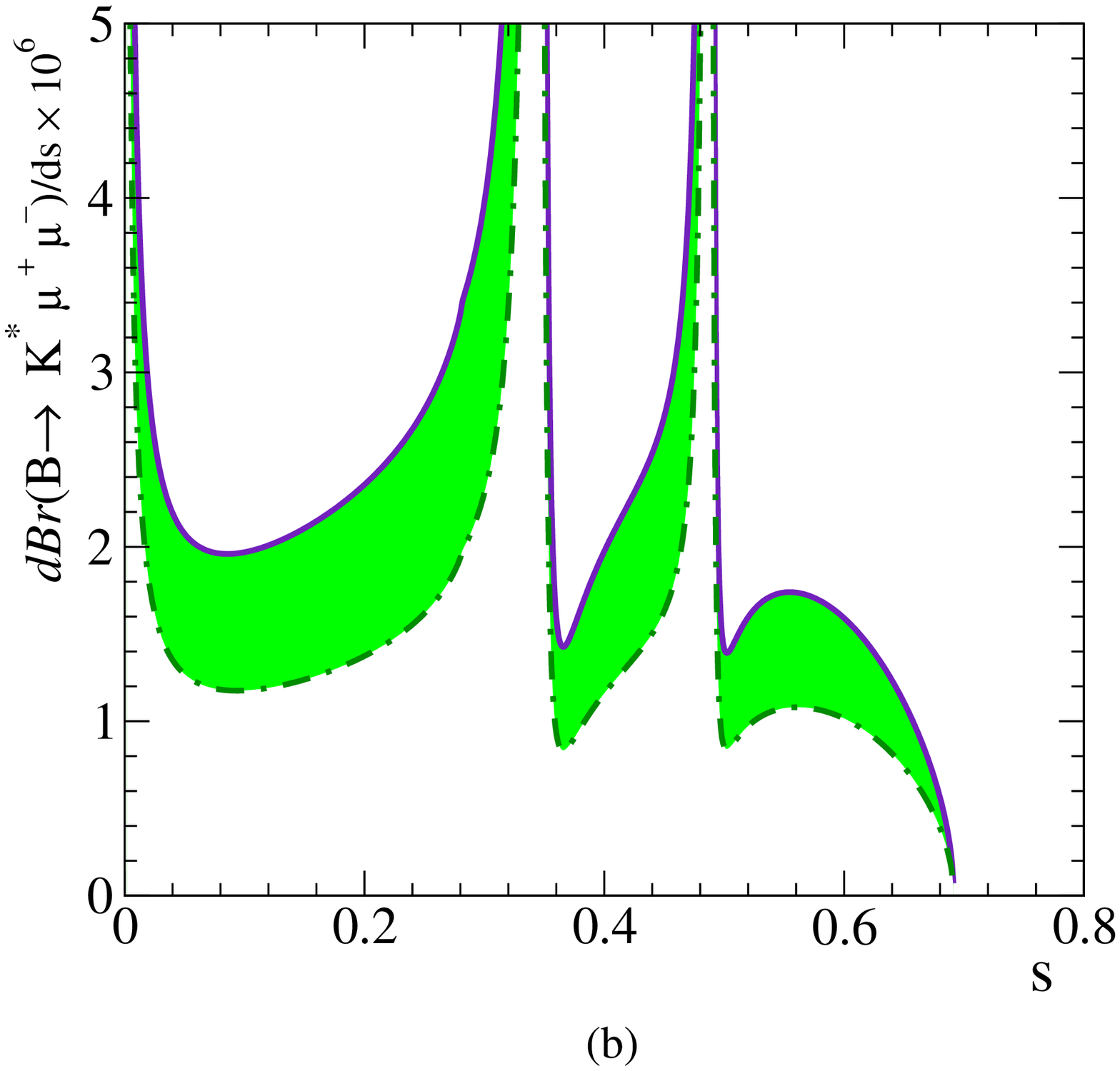,height=3.in } } \caption{Same
as Figure \ref{difratek} but
 $B\rightarrow K^* \mu^+ \mu^+$.}
\label{difrateks}
\end{figure}

In Figures \ref{difratek} and
\ref{difrateks}, we show the differential decay rates of
$B\rightarrow K^{(*)}\mu^+\mu^-$ as functions of $s=q^2/M_B^2$ with and
without resonant $\bar{c} c$ states, respectively.
 From both figures, it is obvious to see that the
dilepton invariant distributions for the two modes
are quite different. However, we expect that the distribution for
$B\rightarrow K \ell^+ \ell^-$, which contains only the
longitudinal part due to the angular momentum conservation, is the
similar to the corresponding part in $B\rightarrow K^* \ell^+
\ell^-$.
For $B\rightarrow K^* \ell^+ \ell^-$,
the transverse part, associated with $1/q^2$ from the
$\gamma$-penguin diagram described by $C_7(\mu)$, gives the
dominant contribution when $q^2$ goes to the allowed minimal values
of $4m_{\ell}^{2}$. This not only
explains why the differential decay rates increase as $q^2\to 0$ but
also indicates the reason for the difference between BRs in
$B\to K^*e^+e^-$ and $B\to K^*\mu^+\mu^-$.

Using Eq. (\ref{difrate}) and the fitted form factors in Tables
\ref{pqcdhqetlattv} and \ref{pqcdhqetlattp}, we obtain
\begin{eqnarray}
Br(B\rightarrow K \ell^+ \ell^-)&=&
(0.53\pm 0.05^{+0.10}_{-0.07}) \times 10^{-6},
\nn\\
Br(B\rightarrow K^* e^+ e^-)&=&
(1.68\pm 0.17^{+0.14}_{-0.09}) \times 10^{-6}, 
\nn\\
Br(B\rightarrow K^* \mu^+ \mu^-)&=&(1.34\pm 0.13^{+0.11}_{-0.06})
\times 10^{-6},
\label{BRv}
\end{eqnarray}
where the first and second errors are from $\lambda_t$ and the
hadronic effects, shown in Tables \ref{pqcdhqetlattp} and
\ref{pqcdhqetlattv}, respectively.
We remark that the central value of Eq. (\ref{BRv}) for
$B\rightarrow K^* e^+ e^-$ is 
the same as the BABAR recent measured value \cite{BaBar1} and it is 
much smaller than that in (III) of Ref. 
\cite{CG-PRD} but that for $B\rightarrow K^* \mu^+ \mu^-$  a little
larger.

\begin{figure}[tbp]
\vspace{0cm} \centerline{ \psfig{figure=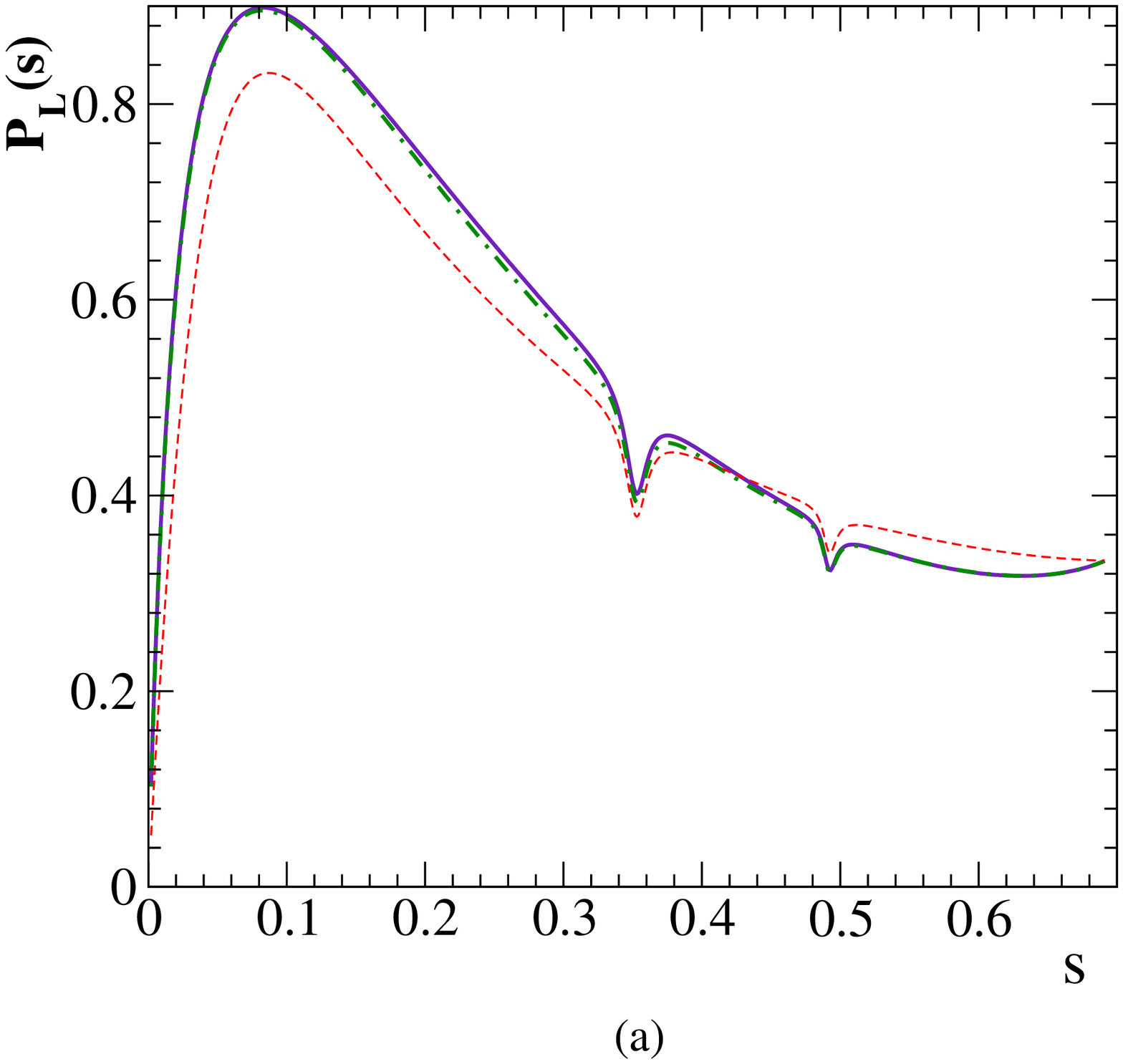,height=3.in } $\ \
$ \psfig{figure=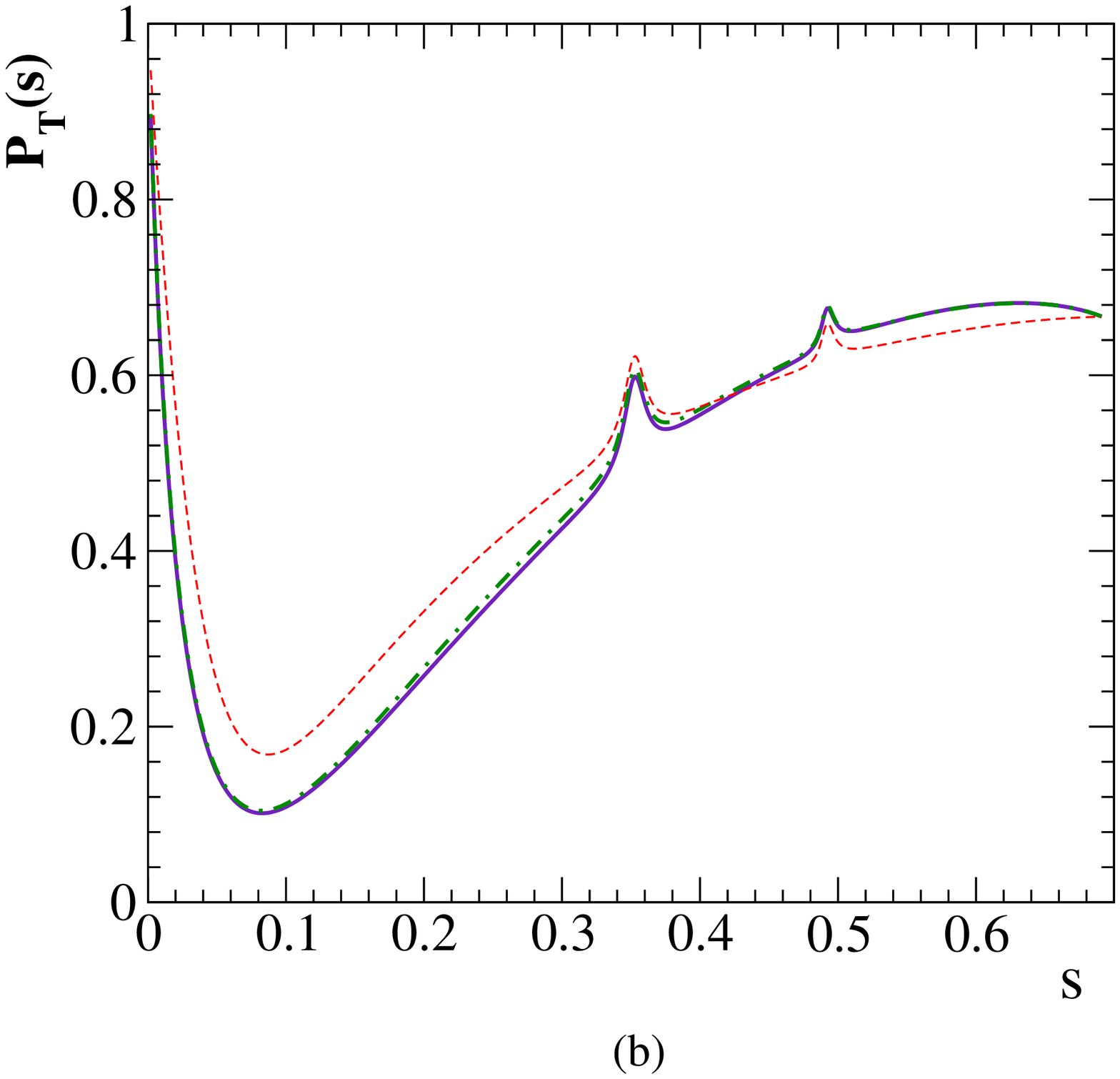,height=3.in } } \caption{Normalized
Longitudinal (a) and transverse (b) polarizations in $B\to
K^*\mu^+\mu^-$. The solid, dash-dotted and dashed curves denote the
results from the PQCD (I), (II) and (III) respectively. } \label{figpl}
\end{figure}

\begin{figure}[tbp]
\vspace{2cm} \centerline{ \psfig{figure=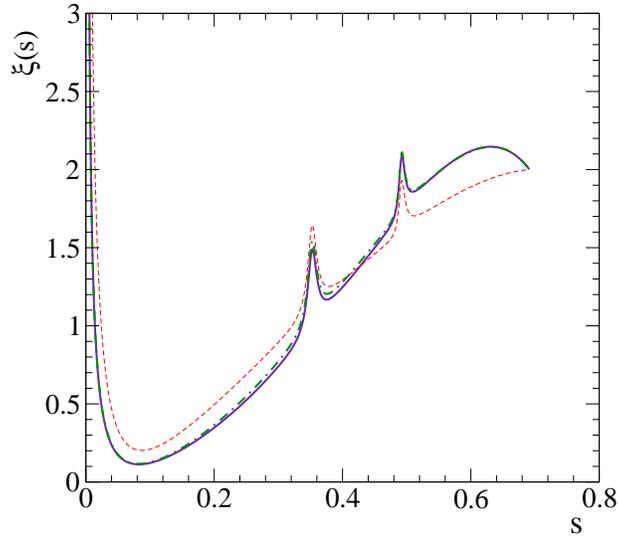,height=3.in }  }
\caption{The ratio of $\xi (s)={\cal P}_{T}(s)/{\cal P}_{L}(s)$ as
a function of $s$. Legend is the same as Figure \ref{figpl}.
}\label{figxi}
\end{figure}

In Figure \ref{figpl}, we present the effects of $K^*$
longitudinal and transverse polarizations based on Eqs. (\ref{pl})
and (\ref{pt}).
In Figure
\ref{figxi}, we display the ratio of $\xi (q^2)$
and we see that when $s\leq 0.016$ ,
where $B\rightarrow K^* \gamma$ is the main effect, and $s \geq
0.339$, where the longitudinal contribution is suppressed, the
transverse part of the $K^*$ polarization becomes dominant.
 We find that both ${\cal P}_{L}$ and ${\cal
P}_{T}$ as well as $\xi (q^2)$
are insensitive to the hadronic uncertainties in 
(I) and (II). 
Moreover, the differences between the results of (I,II) and (III) 
 are small.

\section{T and P violating effects}

\begin{figure}[tbp]
\vspace{0cm} \centerline{ \psfig{figure=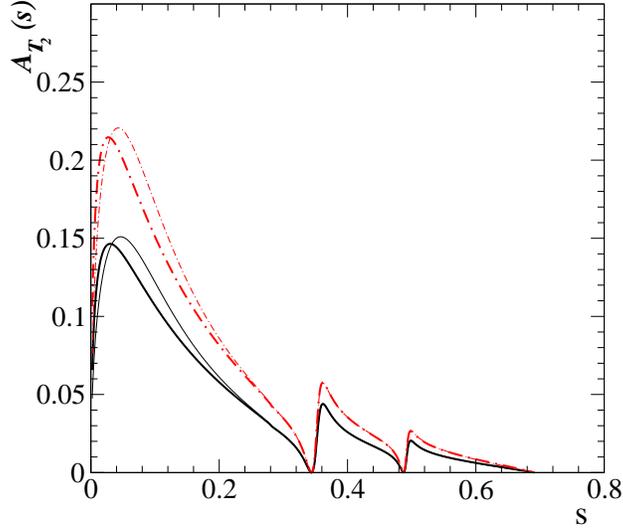,height=3.in } }
\caption{ $T$ violating effect of ${\cal A}_{T_2}(s)$ for (i) 
$ImC_7(\mu)=0.25$ (solid
curves) and (ii) $ImC_7(\mu)=0.25$ and $ImC_{10}=-2.0$ (dash-dotted
curves), where the bold and thin lines correspond to the PQCD (I) and 
(III), respectively.} \label{tv}
\end{figure}

The terms with
 imaginary parts in Eq. (\ref{difangle}) are related to
{\em T odd} effects which, without final state interactions, are
{\em T violating} and thus {\em CP violating} due to the CPT theorem.
In a three-body decay, the triple correlations such as
$\vec{s}_k\cdot
\vec{p}_{i}\times \vec{p}_{j}$ are examples of the effects,
where $\vec{s}_k$ denotes the spin vector carried by one of involving
particles and $\vec{p}_{i,j}$ are the momentum vectors of outgoing
particles.
In the decays of $B\rightarrow K^{*}\ell^{+}\ell^{-}$, the spin
$s_k$ can be either the polarized lepton, $s_{\ell},$ or the
$K^{*} $ meson, $\epsilon ^{*}(\lambda )$. However, since
 the lepton polarization is always associated with the lepton mass
and expected to be
suppressed and less than $1\%$ for the $e$ or $\mu$ mode.
On the other hand, the T odd effects with $K^*$ are free of
the mass suppression and they can be large in models with new physics
\cite{CG-NPB,CG-PRL}. To study these effects, we define
\cite{CG-NPB,CG-PRL}

\begin{eqnarray}
\langle {\cal O}_{i} \rangle=\int {\cal
O}_{i}\omega_{i}(u_{\theta_{K}},u_{\theta_{\ell^+}}){d\Gamma \over
dq^2} \label{op}
\end{eqnarray}
where $\omega_{i}(u_{\theta_{K}},u_{\theta_{\ell^+}})
=u_{\theta_K}u_{\theta_{\ell^+}}/ |u_{\theta_{K}}\,u_{\theta_{\ell^+}}|
$ are sign
functions with $u_{\theta_i}$ being $\cos \theta_i$ or $\sin \theta_i$.
In the $K^*$ rest frame, we use
the  T odd momentum correlations as the operators in Eq. (\ref{op}), given
by
\begin{eqnarray}
{\cal O}_{T_{1}}&=& | \vec{p}_{B}| \frac{(
\vec{p}_{B}\cdot \vec{
p}_{l^{+}}\times \vec{p}_{K}) ( \vec{p}_{B}\times \vec{p}%
_{K}) \cdot ( \vec{p}_{l^{+}}\times \vec{p}_{B}) }{|
\vec{p}_{B}\times \vec{p}_{K}| ^{2}| \vec{p}_{l^{+}}\times \vec{p}%
_{B}| ^{2}} =\frac{1}{2}\sin 2\phi,
\label{OT1}
\\
{\cal O}_{T_{2}}&=& | \vec{p}_{B}|
\frac{\vec{p}_{K}\cdot ( \vec{ p}_{B}\times \vec{p}_{l^{+}}) }{|
\vec{p}_{B}\times \vec{p} _{K}| | \vec{p}_{B}\times
\vec{p}_{l^{+}}| } =\sin \phi,
\label{OT2}
\end{eqnarray}
accompanied with sign functions of
$\omega_{T_{1}}(\sin \theta_{K}, \sin
\theta_{\ell^+})$ and $\omega_{T_{2}}(\cos \theta_{K},
\sin \theta_{\ell^+})$, respectively.
By defining the physical observables as
\be
{\cal A}_i &=&
{\langle {\cal O}_{i} \rangle\over d\Gamma/dq^2} \,,
\label{Ai}
\ee
 from Eqs. (\ref{op}),
(\ref{OT1}) and (\ref{OT2})
 we obtain that
\begin{eqnarray}
{\cal A}_{T_{1}}&=& -{\sum_{i=1,2}Im({\cal M}_{i}^{+}{\cal
M}_{i}^{-*})\over 4 \sum_{\lambda=+,0,- }\sum_{i=1,2}|{\cal
M}_{i}^{\lambda }|^{2} } \label{AT1}
\\
{\cal A}_{T_{2}}&=&\frac{3\pi}{16} {{Im}{\cal
M}_{1}^{0}({\cal M}_{2}^{+*}+ {\cal M}_{2}^{-*})-{Im}({\cal
M}_{1}^{+}+{\cal M}_{1}^{-}){\cal M}_{2}^{0*} \over
\sum_{\lambda=+,0,- }\sum_{i=1,2}|{\cal M}_{i}^{\lambda }|^{2} }
\label{AT2}
\end{eqnarray}
We note that as shown in Refs. \cite{CG-NPB,CG-PRL} ${\cal
A}_{T_1}$ and ${\cal A}_{T_2}$ depend on
${Im}C_{9}^{eff} (\mu )C_{7}(\mu )^*$ and ${Im}C_{7}(\mu
)C_{10}^{*}$, respectively. In the SM,  ${\cal A}_{T_2}$ is zero
since
there are no absorptive parts expected from $C_{7}$ and $C_{10}$,
whereas ${\cal A}_{T_1}$ is non-zero  due to that from
$C_{9}^{eff}$. However, it is clear that the T-odd observable of
${\cal A}_{T_2}$ can be large in models with new CP
violating phases beyond the SM.
To illustrate a new physics result, in Figure \ref{tv} we show the T
violating  effect of ${\cal A}_{T_2}$ as a function of $s$ by
taking two cases with the imaginary parts of WCs as follows: (i)
$ImC_7(\mu)=0.25$ and
 (ii) $ImC_7(\mu)=0.25$ and $ImC_{10}=-2.0$.
One possible origin of having these imaginary parts is from SUSY
\cite{CG-PRL}
 where there are many CP violating sources.
Here, we only show the cases with (I) and (III) while those in (II) are
almost the same as (I).
It is interesting to note that the CP violating effect can be as
large as $15\%$ in both cases as shown in Figure \ref{tv}.
We remark that
 ${\cal A}_{T_1}$ is much
smaller than ${\cal A}_{T_2}$ in most of cases with new
physics.

 From Eq. (\ref{difangle}), we can also study another
interesting physical observable associated with the angular
distribution of $\sin 2\theta_{K} \cos \phi$ in $B\to K^* \ell^+
\ell^-\to K\pi \ell^+ \ell^-$.
This observable is defined as an up-down asymmetry (UDA)
of the $K$ meson due to its dependence of $\cos \theta_{K}$.
Explicitly, we define that
\begin{eqnarray}
{\cal O}_{UD}&=&{(\vec{p}_{B}\times \vec{p}_{K})\cdot
(\vec{p}_{\ell^+}\times\vec{p}_{B}) \over |\vec{p}_{B}\times
\vec{p}_{K}||\vec{p}_{\ell^+}\times\vec{p}_{B}|}=\cos\phi\,,
\nonumber \\
\omega(u_{\theta_{K}},u_{\theta_{\ell}})&=&\omega(\cos\theta_{K},
\sin\theta_{\ell})\,,
\end{eqnarray}
as the operator
corresponding to the UDA of $K$
in the $K^*\ (\to K\pi)$ rest frame
and from Eqs. (\ref{difangle}), (\ref{op}) and (\ref{Ai})
we find that
\begin{eqnarray}
{\cal A}^{K}_{UD}(q^2)\equiv {\langle {\cal O}_{UD} \rangle \over
d\Gamma/dq^2}={3\pi \over 16}{ {Re}{\cal M} _{1}^{0}({\cal
M}_{2}^{+*}-{\cal M}_{2}^{-*})+{Re}({\cal M}_{1}^{+}-{\cal M}
_{1}^{-}){\cal M}_{2}^{0*} \over  \sum_{\lambda=+,0,-
}\sum_{i=1,2}|{\cal M}_{i}^{\lambda }|^{2} }\,,
 \label{fbaks}
\end{eqnarray}
which clearly violates parity.

\begin{figure}[tbp]
\vspace{2cm} \centerline{ \psfig{figure=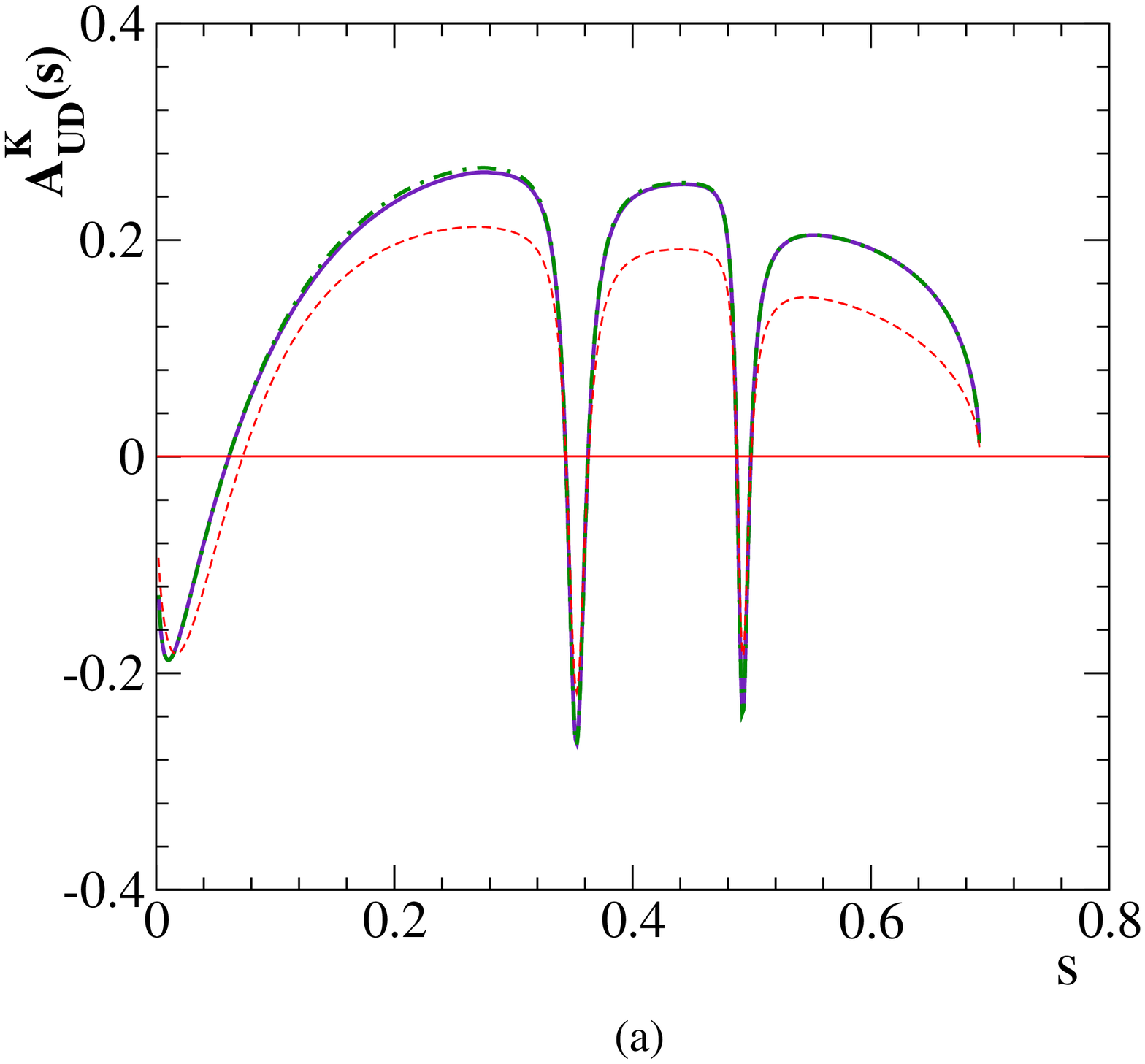,height=3.in }
\psfig{figure=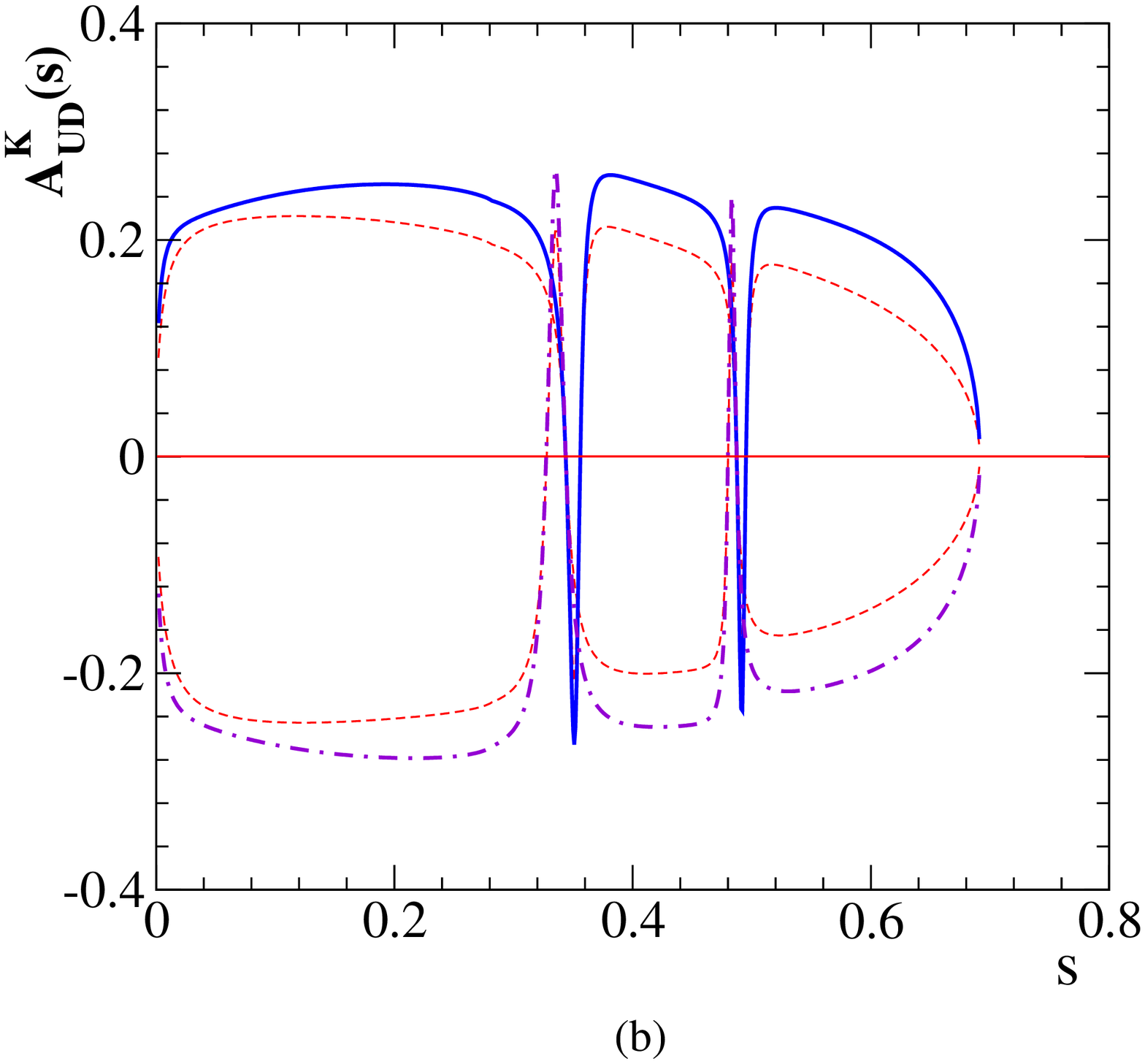,height=3.in }} \caption{${\cal
A}_{UD}^K(s)$ as a function of $s$. In (a) the solid,
dash-dotted and dashed curves represent the SM contributions based on the
form factors in the PQCD (I), (II) and (III), while
in (b) the solid (dash-dotted) and upper (lower) dashed curves
are for $C_{7}(\mu)>0\ (C_{9}(\mu)<0)$ in (I) and (III),
 respectively.}
\label{figfbaks}
\end{figure}

In Figure \ref{figfbaks}a, we show the $ {\cal A}^{K}_{UD}(s)$ as
a function of $s=q^2/M_B^2$ based on the form factors given by the
PQCD (I), (II) and (III) in the SM. It is interesting to point out that,
as shown in the figure, in the SM ${\cal A}^{K}_{UD}(s)$ crosses
zero point at $s_0\simeq 0.08$ which satisfies the identity
\begin{eqnarray}
Re\Big(h_{1}g^{*}_{2}+h_{2}g^{*}_{1}
\Big)=-2{\sqrt{m^{2}_{B}s_{0}}|\vec{p}_{K^*}|^{2}\over
E_{K^*}}Re\Big( h_{3}g^{*}_{1}+h_{1}g^{*}_{3}\Big)\,.
\end{eqnarray}
We note that the point $s_0$ is insensitive the QCD models but the
Wilson coefficients
of  $C_7(\mu)$ and $C_9(\mu)$, especially the relative sign between
them.
To show the result, in Figure \ref{figfbaks}b we present
 two extreme cases of
(i) $C_7(\mu)=-C_7(\mu)_{SM}$ (solid curve) and
(ii) $C_9(\mu)=-C_9(\mu)_{SM}$
(dash-dotted curve)
with  the remaining Wilson coefficients the same as those in the SM,
respectively. From the figure, we see that
the distributions in the two cases are quite different and moreover,
 $s_0$ disappears,
 $i.e.$, it exists only if  $C_7(\mu)$ and $C_9(\mu)$
have an opposite sign since $C_7(\mu)_{SM}<0$ and $C_9(\mu)_{SM}>0$.
It is clear that $s_0$
provides us a good candidate to explore new physics
due to the insensitivity of the QCD models and dependence on the WCs.

\section{Conclusions}

We have studied the exclusive decays of $B\to K^{(*)}\ell^+
\ell^-$ by the results in the PQCD with the HQET and LQCD. We have
given the form factors for the $B\to K^{(*)}$ transitions in the
whole allowed region, which are consistent with those from other
QCD models. We have found that the branching ratios of $B\to K
\ell^{+} \ell^{-}$, $B\to K^{*} e^{+}e^{-}$, and $B\to K^{*}
\mu^{+} \mu^{-}$ are $0.53\pm 0.05^{+0.10}_{-0.07}$, $1.68\pm 0.17
^{+0.14}_{-0.09}$, and $1.34\pm0.13^{+0.11}_{-0.06}\times
10^{-6}$, respectively. We have shown that, in $B\to K^*\ell^+
\ell^-\to K\pi\ell^+ \ell^-$, the T-odd observable of ${\cal
A}_{T2}$ which is unmeasurably small in the SM could be as large
as $20\%$ in models with new physics, while
 the P-odd up-down asymmetry of ${\cal A}_{UD}^K(s)$ vanishes at
$s_0 \simeq 0.08$ in the SM but it behaves quite differently if
new physics exists, which provide
us unique probes of non-standard physics.\\


 \noindent {\bf Acknowledgments}

This work was supported in part by
 the National Science Council of the Republic of China under
 Contract Nos. NSC-90-2112-M-001-069 and NSC-90-2112-M-007-040 and
 the National Center for Theoretical Science.

\newpage
\begin{center}
{\bf\large Appendix}
\end{center}

The parameters of $h_{1,2,3}$ and $g_{1,2,3}$ in Eq. (\ref{ampk1})
are defined by
\begin{eqnarray*}
h_{1} &=&C_{9}(\mu){V(q^{2})\over m_{B}+m_{K^*}}
+\frac{2m_{b}}{q^{2}}(\mu
)C_{7}(\mu)T_{1}(q^{2}), \nonumber \\
h_{2} &=&-C_{9}(\mu
)(m_{B}+m_{K^*})A_{1}(q^{2})-\frac{2m_{b}}{q^{2}} P\cdot q
C_{7}(\mu)T_{2} (q^{2}),  \nonumber \\
h_{3} &=&C_{9}(\mu ){A_{2}(q^{2})\over
m_{B}+m_{K^*}}+\frac{2m_{b}}{q^{2}} C_{7}\Big(\mu)(T_{2}
(q^{2})+\frac{q^2}{P\cdot q}T_{3}(q^2)\Big),
\nonumber \\ g_{1} &=&C_{10}{V(q^{2})\over m_{B}+m_{K^*}}, \nonumber \\
g_{2} &=&-C_{10}(m_{B}+m_{K^*})A_{1}(q^{2})\,, \nonumber \\
g_{3} &=&C_{10}{A_{2}(q^{2}) \over m_{B}+m_{K^*}}.
\end{eqnarray*}


\newpage


\end{document}